\documentclass{aa}
\usepackage{graphicx}
\usepackage{txfonts}
\begin{document}
   \title{Noise Properties of the Planck-LFI Receivers}

   \author{P. Meinhold\inst{1}, R. Leonardi\inst{1}, B. Aja\inst{2}, E. Artal\inst{2}, P. Battaglia\inst{3},
   M. Bersanelli\inst{4}, E. Blackhurst\inst{5}, C. R. Butler\inst{6}, L. P. Cuevas\inst{7}, F. Cuttaia\inst{6}, O. D'Arcangelo\inst{8},
   R. Davis\inst{5}, M. L. de la Fuente\inst{2}, M. Frailis\inst{9}, C. Franceschet\inst{3}, E. Franceschi\inst{6}, T. Gaier\inst{10}, S. Galeotta\inst{9}, A. Gregorio\inst{11,9},
   R. Hoyland \inst{12}, N. Hughes\inst{13}, P. Jukkala\inst{13}, D. Kettle\inst{14}, M. Laaninen\inst{15}, P. Leutenegger\inst{3}, S. R. Lowe\inst{5},
   M. Malaspina\inst{6}, R. Mandolesi\inst{6}, M. Maris\inst{9}, E. Mart\'{\i}nez-Gonz\'alez\inst{16}, L. Mendes\inst{17}, A. Mennella\inst{4}, M. Miccolis\inst{3},
   G. Morgante\inst{6}, N. Roddis\inst{5}, M. Sandri\inst{6}, M. Seiffert\inst{10}, M. Salm\'on \inst{16}, L. Stringhetti\inst{6}, T. Poutanen\inst{18,19,20},
   L. Terenzi\inst{6}, M. Tomasi\inst{4}, J. Tuovinen\inst{21}, J. Varis\inst{21}, L. Valenziano\inst{6}, F. Villa\inst{6}, A. Wilkinson\inst{5},
   F. Winder\inst{5}, A. Zacchei\inst{9}, and A. Zonca\inst{22}}

   \institute{Department of Physics, University of California, Santa Barbara, CA 93106, USA. \\
              \email{peterm@cfi.ucsb.edu} 
              \and
              Departamento de Ingenieria de Comunicaciones, Universidad de Cantabria, Santander,
              Spain. 
              \and
              Thales Alenia Space Italia S.p.A., IUEL - Scientific Instruments, S.S. Padana Superiore 290, 20090 Vimodrone (Mi),
              Italy. 
              \and
              Universit\`{a} degli Studi di Milano, Dipartamento di Fisica, via Celoria 16, 20133, Milano,
              Italy.
              \and
              Jodrell Bank Centre for Astrophysics, Alan Turing Building, The University of Manchester, Manchester, M13 9PL, UK. 
              \and
              INAF/IASF, via P. Gobetti 101, I-40129, Bologna, Italy.
              \and
              ESTEC, Keplerlaan 1, Postbus 299 2200 AG Noordwijk, Netherlands.
              \and
              Istituto di Fisica del Plasma CNR, via Cozzi 53, 20125 Milan, Italy. 
              \and
              INAF/OATs, via Tiepolo, 11 Trieste, I-34143, Italy. 
              \and
              Jet Propulsion Laboratory, Pasadena, CA 91109, USA. 
              \and
              University of Trieste, Department of Physics, via Valerio, 2 Trieste I-34127,
              Italy.
              \and
              Instituto de Astrof\'{i}sica de Canarias, C/ V\'{i}a L\'{a}ctea S/N, E-38200, La Laguna, Tenerife,
              Spain. 
              \and
              DA-Design Oy, Keskuskatu 29, FI-31600 Jokioinen, Finland. 
              \and
              School of Electrical and Electronic Engineering, The University of Manchester, Manchester, M60 1QD,
              UK. 
              \and
              Ylinen Electronics Oy, Teollisuustie 9, FI-02700 Kauniainen, Finland. 
              \and
              Instituto de F\'{\i}sica de Cantabria, CSIC-Universidad de Cantabria, Avenida Los Castros s/n, 39005 Santander,
              Spain. 
              \and
              Planck Science Office, European Space Agency ESAC, P.O. box 78 28691 Villanueva de la Ca\~{n}ada Madrid,
              Spain.
              \and
              University of Helsinki, Department of Physics, P.O. Box 64, FI-00014 Helsinki,
              Finland. 
              \and
              Helsinki Institute of Physics, P.O. Box 64, FI-00014 Helsinki, Finland. 
              \and
              Mets\"ahovi Radio Observatory, Helsinki University of Technology, Mets\"ahovintie 114, FI-02540 Kylm\"al\"a, Finland. 
              \and
              MilliLab, VTT Technical Research Centre of Finland,
              P.O. Box 1000, FI-02044 VTT, Finland. 
              \and
              INAF/IASF Milano, Via Bassini, 15, 20133, Milano, Italy.
             }



  \abstract
{}
{The Planck Low Frequency Instrument (LFI) radiometers have been
tested extensively during several dedicated campaigns. The present
paper reports the principal noise properties of the LFI
radiometers.}
{A brief description of the LFI radiometers is given along with
details of the test campaigns relevant to determination of noise
properties.}
{Current estimates of flight sensitivities, $1/f$ parameters, and
noise effective bandwidths are presented.}
{The LFI receivers exhibit
exceptional $1/f$ noise, and their white noise performance is sufficient for the science goals of Planck.}


   \authorrunning{P. Meinhold et al.}

   \maketitle

\section{Introduction}

The Low Frequency Instrument (LFI), installed on board the
European Space Agency's Planck satellite, is designed to measure
temperature and polarization anisotropies of the cosmic microwave
background (CMB) in three frequency bands from 30 to 70 GHz. The
core of the Planck-LFI is a compact Radiometer Array Assembly
(RAA) of 22 pseudo correlation radiometers, with cryogenic
low-noise microwave amplifiers, which are coupled to the 1.5 meter
Planck telescope by an array of 11 conical dual profiled
corrugated feed horns. Design, construction and testing of the LFI
are extensively described in a set of accompanying papers: Mandolesi (\cite{mandolesi09}), Bersanelli (\cite{bersanelli09}), Mennella (\cite{mennella09a}), Villa (\cite{villa09}), Sandri (\cite{sandri09}), Leahy (\cite{leahy09}). This work reports the noise
performance of the Planck-LFI receivers measured in ground tests,
and expected in-flight sensitivity.

In Section \ref{SectionOverview}, we present a brief overview of the LFI receivers and their
data acquisition system. In Section \ref{SectionNoiseModel}, we outline the noise properties
being investigated. In Section \ref{SectionTestCampaign}, we describe the test campaigns
and the main results obtained from them. In Section \ref{SectionSummary}, we provide a summary and
discussion of the LFI noise performance. In Section \ref{SectionConclusion}, we present
the conclusion of this work.

\section{Overview of the LFI receivers and data acquisition}\label{SectionOverview}

A key objective of the Planck-LFI radiometer architecture is
minimizing $1/f$ noise. Excess $1/f$ noise would degrade the
radiometer's sensitivity, increase the uncertainty in the measured
angular power spectrum at low-$\ell$, and add a source of
systematic errors that would propagate in a non-trivial way
through the Planck-LFI scientific products (e.g. Maino \cite{maino02}).

Planck will observe the sky by continuously scanning nearly great
circles on the celestial sphere with a one minute period, and
periodically (approximately 50 minutes) shifting the spin axis to
remain anti-sun throughout the year.

Each of the 11 LFI feed horns couples radiation from the Planck's
optics through a Receiver Chain Assembly (RCA), that consists
of an actively-cooled 20 K Front-End Module (FEM) connected via
waveguides to a 300 K Back-End Module (BEM), which is followed by
the Data Acquisition Electronics (DAE), and Radiometer Electronics
Box Assembly (REBA).

An Orthomode Transducer (OMT) separates the radiation that enters
the RCA FEM into two orthogonally polarized components, and
transmits each component to a pseudo correlation radiometer. A
$180^\circ$ hybrid coupler combines the sky signal with the signal
from a cooled (approximately 4 K) reference load viewed with a
small feed horn. The two outputs of the hybrid are then amplified
by cryogenic low-noise High Electron Mobility Transistor (HEMT)
amplifiers and each passing through a 180 degree phase switch. One of the phase
switches is modulated at 4096 Hz whilst the other remains unswitched. The two resulting signals run into
a second 180$^\circ$ hybrid coupler. The FEM RF signals are then
transmitted via waveguides to the Back End Module (BEM), where
they are further amplified, filtered, and detected by square-law
diodes (Aja \cite{aja05a}, \cite{aja05b}, Artal \cite{artal09}). The BEMs include preamplifiers 
the detector diodes to raise the signal level before input to the Data Acquisition Electronics (DAE). The outputs of the detectors form two
independent streams of data alternating between sky and reference
at 4096 Hz for each radiometer, or four independent streams per RCA. A general
analytical description of the Planck-LFI radiometers is given by
Seiffert (\cite{seiffert02}), and Mennella (\cite{mennella03}).

\begin{figure}
\centering
\includegraphics[width=9cm]{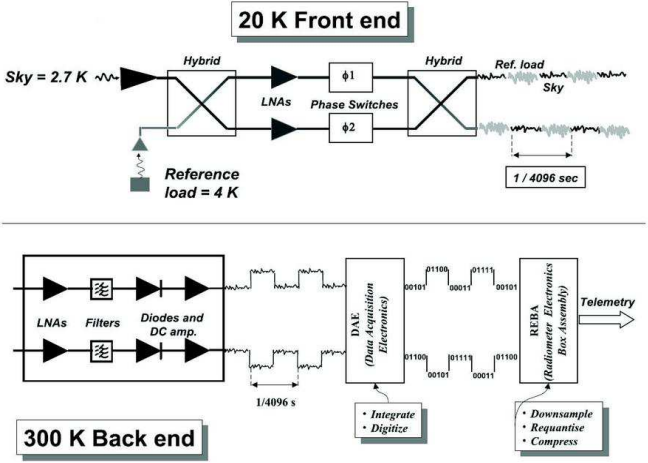}
\caption{Schematic of one LFI radiometer and its data acquisition system. The outputs of the radiometer form two
independent streams of data alternating between the signals from the sky and from a stable internal cryogenic reference load. The radiometer architecture is designed to efficiently
minimize the $1/f$ noise on the post detection differenced time stream defined in Equation \ref{e:equation_diff}. This figure is adapted from Seiffert (\cite{seiffert02}).} \label{LFISchematic}
\end{figure}

The BEM outputs pass to the DAE, where a programmable offset and gain are applied to optimize
the dynamic range. The signals are then integrated in an ideal integrator circuit, and converted
from analog to digital signals. The digitized output is
downsampled, mixed, re-digitized, compressed, and transmitted to telemetry by software in the REBA (Maris \cite{maris09}).
The difference between sky and reference is not calculated at this stage. It is calculated on the ground, after data have been decompressed, demixed and time streams are reconstructed (Zacchei \cite{zacchei09}). With only the data acquired from a given diode we create
sky-only, and reference-only time streams, allowing us to compute
the uncalibrated differenced time stream $V_{\mathrm{diff}}$ as
\begin{equation}\label{e:equation_diff}
    V_{\mathrm{diff}} = V_{\mathrm{sky}}-rV_{\mathrm{ref}},
\end{equation}
where $V_{\mathrm{sky}}$ is the uncalibrated sky-only time stream,
$V_{\mathrm{ref}}$ is the uncalibrated reference-only time stream,
and $r$ is a gain modulation factor that brings the difference as
close as possible to zero, simultaneously minimizing $1/f$ and
equalizing the white noise contributions of sky and reference data
samples. A value of $r$ equal to the ratio of the means of the sky and reference time streams achieves this goal (e.g. Mennella \cite{mennella03}; see also discussion in Section \ref{SectionGMFdiscussion}). Finally, the calibrated differenced time stream
$T_{\mathrm{diff}}$ is then computed as
\begin{equation}\label{e:equation_diff_calibrated}
    T_{\mathrm{diff}} = GV_{\mathrm{diff}},
\end{equation}
where $G$ is the photometric calibration, valid in the linear
response range of the instrument, which converts the output
voltage into Rayleigh-Jeans temperature units. Figure \ref{LFISchematic} displays a schematic of one LFI radiometer and its data acquisition system.

The main advantages of the LFI radiometer design are: (i) its
sensitivity does not depend (to first order) on the absolute level of the
reference signal, even in the case of a slight imbalance of the
radiometer's components; (ii) with the proper gain modulation
factor, the $1/f$ noise in the radiometer time streams depends
mostly on the noise temperature fluctuations of the FEM amplifiers; (iii) the
$1/f$ fluctuations from gain fluctuation in the BEM amplifiers are
effectively removed by the fast switching between sky and
reference signals.

It is worth noting explicitly that the noise performance of LFI, in particular
the $1/f$ noise, is dependent to some extent on every element in the receiver
chain, from the emissivity and temperature stability of the reflectors, to the
frequency dependent match and loss of the passive front end components, the
stability and match of the reference horns and loads, the thermal stability of
the back ends, as well as the electrical and thermal stability of the
downstream electronics. First order fluctuations induced after the FEM are
very effectively removed by the differencing of sky and reference time streams
for each diode, switching at 4 KHz. Second order fluctuations, and those
induced by temperature changes in the front end are not removed by the switching.

As a unique feature of LFI, the 4K reference loads, as well as the reference horns used to view them, have undergone careful design and testing (Valenziano, 2009). The horn/load coupling return loss was measured via network analyzer, and all horn/load combinatgions met the requirement of -20 dB integratedas measured by the Return Loss reference horn-loads ( using SNA on optical bench)
The requirement was to be better then -20 dB

All the reference loads satisfy this requirement; performance change with the frequency , since loads and horns have been projected different depending on the frequency ( actually waveguides dimensions do not scale in size with frequency , as well as the gaps reference horn - reference load, being constant while Lambda is changing with channels) .
The req. is given as integrated mismatching along the full bandwidth , hence the return loss level for particular frequencies can also be higher than the requirement .
The reference horn insertion loss was roughly measured by shortening the horn and measuring S11: result is a worst case and is in all the cases better than 0.15 dB .
The spillover, accounting for the internal straylight , is lower than -40 dB ; external sky radiation, passing through the main frame , is further dumped and is found lower than -60 dB . In both cases spillover is just simulated and not calculated.

In this paper we analyze results of tests with representative conditions
achievable in the instrument test campaigns.  In particular, we do not describe data where the flight
cooling chains were employed. The test campaigns included explicit tests for
susceptibility to the various effects noted above but the actual fluctuations induced by the
cooling chains were not measured until the satellite integration tests (not
covered by this work). The impact of the sorption
cooler and 4 K cooler on the performance of LFI is non trivial, but has been
investigated through extensive simulations and measurements of the cooling
chain. There is a specification for fluctuations of the 4 K load of 10
$\mu\mathrm{K}/\sqrt{\mathrm{Hz}}$. In general, $1/f$ is dominated by the receiver
characteristics, not random thermal fluctuations in the loads or passive
optical elements. We do anticipate measurable signals in LFI due to the
cycling of the sorption cooler. This is regular and predictable, and couples to
the radiometer outputs through the channels mentioned above. These issues are
explored most comprehensively in Terenzi \cite{terenzi09a}, and \cite{terenzi09b}. All of these effects will be
studied exhaustively both using the satellite test campaign, and of course the
flight data.

\section{Noise model}\label{SectionNoiseModel}

Several tests were performed on the RCA and RAA instruments to
investigate and characterize the noise of the LFI radiometers.

We parameterize the noise power spectral density $P(f)$ as
\begin{equation}\label{e:PSD}
    P(f)\sim\sigma^{2}\left[1+\left(\frac{f}{f_{\mathrm{k}}}\right)^{\alpha}\right],
\end{equation}
where $\sigma$ characterizes the white noise component, the knee
frequency $f_{\mathrm{k}}$ denotes the frequency where white noise
and $1/f$ contribute equally to the total noise, and $\alpha$
characterizes the slope of the power spectrum for frequencies
$f<f_{\mathrm{k}}$.

We use the radiometer equation to evaluate how well the white
noise level corresponds to expectation based on the system
temperature, which is given as
\begin{equation}\label{e:radiometereq}
\sigma_{T}=K\left(\frac{T_{\mathrm{noise}}+T_{\mathrm{target}}}{\sqrt{\beta_{\mathrm{eff}}\tau}}\right),
\end{equation}
where $\sigma_{T}$ is the root-mean-square noise,
$T_{\mathrm{noise}}$ is the system noise temperature,
$T_{\mathrm{target}}$ is the antenna temperature of the target
being observed, $\beta_{\mathrm{eff}}$ is the noise effective
bandwidth, $\tau$ is the integration time, and $K$ is a constant
of order unity which depend on the receiver architecture (see
Appendix A for details), e.g. Kraus (\cite{kraus86}).

In addition to tracking the white noise level, knee frequency and
slope, we use the noise effective bandwidth $\beta_{\mathrm{eff}}$
as a figure of merit for checking white noise consistency. The noise
effective bandwidth can be interpreted as the width of an ideal
rectangular RF band-pass filter which would produce the same noise
power as the noise measured from the real instrument. In an ideal radiometer (i.e. when input target temperature is proportional to output voltage signal), it can be
estimated from the data as
\begin{equation}\label{e:NEB}
\beta_{\mathrm{eff}}=\left(\frac{KV_{\mathrm{DC}}}{\sigma_{\mathrm{V}}}\right)^{2},
\end{equation}
where $V_{\mathrm{DC}}$ is the uncalibrated offset when observing
the target, and $\sigma_{\mathrm{V}}$ is the uncalibrated white
noise. The noise effective bandwidth is independent of gain,
independent of system and target temperatures, specifically
sensitive to voltage offsets, and external noise such as
electromagnetic interference (EMI). This is why
$\beta_{\mathrm{eff}}$ is a good figure of merit for checking
noise consistency. It should be noted here that because the RAA test campaign operated the 30 and
44 GHz radiometers in a slightly compressed state, a minor modification to Equation \ref{e:NEB} is needed for these receivers, and $\beta_{\mathrm{eff}}$ should be estimated as
\begin{equation}\label{e:NEBcompressed}
\beta_{\mathrm{eff}}=\left(\frac{KV_{\mathrm{DC}}}{\sigma_{\mathrm{V}}}\right)^{2}\times[1+bG(T_{\mathrm{noise}}+T_{\mathrm{target}})]^{-2},
\end{equation}
where $T_{\mathrm{noise}}$ is the system noise
temperature, $T_{\mathrm{target}}$ is an input temperature, and $b$ is
a non linearity parameter.
This is discussed in detail in Mennella (\cite{mennella09a}, \cite{mennella09b}).

Investigation of noise properties of LFI during the several test campaigns
characterized the instrument performance, but also served to uncover subtle
systematic effects from
instrument anomalies.
In all cases the analysis allowed us to track down and solve the problems.
One example was the occurrence of `telegraph' or `popcorn' noise, which was
traced to an oscillating BEM which was eventually replaced. During
another campaign, excessive $1/f$ noise in a 44 GHz radiometer was traced to a damaged phase
switch on a 70 GHz radiometer which shares
the same power supply. This was solved by changing the operating mode of
the 70 GHz phase switch.
During test campaigns in Finland and Laben several frequency spikes were seen
in the power spectra, which were finally proven to be from the ground
electronics, and completely absent in the satellite level tests. An exception
is the set of spikes caused by the housekeeping data acquisition, which is
described in detail
in section \ref{SectionFrequencySpikes}. Ultimately,
the best description of the LFI noise properties is the very
simple model given by Equation \ref{e:PSD}.

\section{Test campaign}\label{SectionTestCampaign}

The LFI test campaign included breadboard, qualification model,
and flight model tests. Testing was performed at amplifier level,
RCA level, and RAA level (Tanskanen \cite{tanskanen00}, Kangaslahti \cite{kangaslahti01}, Sj\"{o}man \cite{sjoman03}, Laaninen \cite{laaninen06}, Terenzi \cite{terenzi09a}, Davis \cite{davis09}, Artal \cite{artal09}, Villa \cite{villa09}, Varis \cite{varis09}). This work focuses on the
flight model tests at RCA and RAA level. Noise properties
considered here were measured after an extensive tuning procedure (Cuttaia \cite{cuttaia09a}),
which involved optimizing the cryogenic amplifier biases for
system noise and bandwidth, matching phase switch response, and
tuning for overall `isolation' between the two output states which
measure sky and reference load temperatures.

\subsection{RCA Campaigns}

Basic parameters for the LFI receivers were measured in detail
during the RCA test campaigns. These measurements and results are
discussed in Villa \cite{villa09} and references above. In addition we emphasize the following relevant
points about these data sets.
\begin{enumerate}
\item Direct swept source measurements of the bandwidths are
generally consistent with noise estimated bandwidths (Zonca \cite{zonca09}). \item Measured temperature
calibrated white noise is generally consistent with the measured
$T_{\mathrm{noise}}$ and Equation \ref{e:radiometereq}. \item For 30 and 44 GHz RCAs, the measured $1/f$ knees
are well below the specification of 50 mHz, while for 70 GHz the
sky load was not stable enough to measure this. \item Gain
compression was measured for all RCAs. All will be in the linear
regime for flight operations. \item 70 GHz RCAs are linear over
all test conditions (RCA and RAA). \item 30 and 44 GHz RCAs are
compressed. Thus careful calibration of the compression curves were carried
out to help predict performance for different target temperatures.
\end{enumerate}

\subsection{RAA Campaign}

After RCA testing, the receivers were installed in the full array and retested as the RAA in the Thales Alenia Space Italia laboratories. In this test flight hardware (electronics, harnesses and computer) was used. This test campaign
included tuning, interference tests, system temperature, and noise
characterization among many other things. A single large sky load
was used, with very good long term stability. This allowed more
careful measurement of LFI's very low $1/f$ noise. For detailed
description of the test systems, including thermal, electrical, RF
and mechanical design of the loads and chamber see Terenzi (\cite{terenzi09b}), Morgante (\cite{morgante09}), and Cuttaia (\cite{cuttaia09b}).

Due to the size of the sky load and the design of the chamber, RAA
sky and reference loads could not be cooled below about 18 K, and
sky load time constants were many hours. This allowed complete
system tests, $1/f$ measurements, crosstalk measurements, tuning
etc, but provided some limitations with respect to noise
parameters.

System temperature measurements from RCA tests were more reliable
than from RAA tests, due to complications from gain compression,
thermal gradients and limited temperature step sizes. The long time constants of the loads for RAA
testing also limited the number of temperature steps to three, leaving the fits poorly constrained.

Temperature calibrated white noise levels were similarly affected
by compression and temperature gradients (causing systematic
errors in our estimates of the gain). Noise temperature and, ultimately, white noise sensitivity were also affected by the fact that the Focal Plane Unit (FPU) temperature was kept at about 26 K, instead of the 20 K as it is going to be the case during flight.

A campaign to analyze the thermal and RF properties of
the full system has been carried out with some success, and is
described in some of the references above. However, for our purposes in characterizing the
LFI FM noise performance we prefer to take the best measurements
from each of the test campaigns, in addition to providing evidence
for consistency among tests wherever possible.

\subsubsection{RAA results}\label{SectionRAAResults}

\begin{figure}
\centering
\includegraphics[width=9cm]{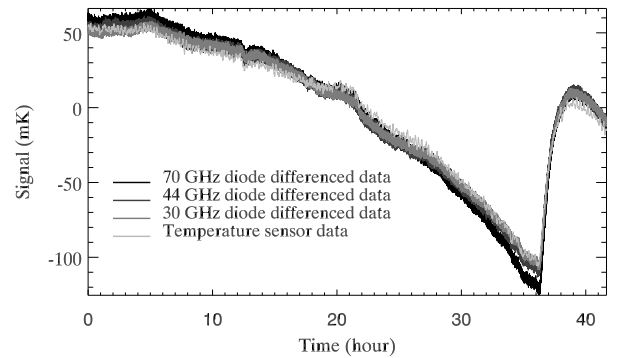}
\caption{Comparison between scientific and thermal environment
time streams from data in ST1\_0002. The mean was removed from the
temperature sensor. The radiometer's differenced data follow the thermal
environment behavior. The small differences between them are due
to a temperature gradient in the microwave absorber. The long term
offset changes due to a slow and continuous change in reference
temperature ($\sim 0.7$ mK/hour), which reaches a minimum at 36h.
This behavior was due to a small leak in one of the gaseous helium
heat switches of the cryo chamber. Despite this drift, the test
provided many hours of stable data that were useful for
characterization of LFI RAA noise properties.} \label{FigTod}
\end{figure}

Our primary source for information on the $1/f$ performance of LFI RAA
is a long ($44.3$ hours) data set acquired in near nominal thermal
environment and nominal optimized biases for all channels. During its data acquisition, the physical temperatures of the microwave absorber and the reference load were kept at about 22 K and 19 K, respectively. The
length of the data set allowed  us to investigate stability of the
noise parameters as well as different techniques and timescales
for calculating the gain modulation factor. In addition we have
very good statistics to look for systematic effects such as
crosstalk and anomalous frequency spikes in the data. Data were
acquired with sample rates of 32.5 Hz,
46.5 Hz, and 78.8 Hz, for 30 GHz, 44 GHz, and 70 GHz channels
respectively. Figure \ref{FigTod} shows examples of differenced time streams
from the long data set. For the purposes of noise analysis, the
salient features of the RAA data sets include:

\begin{enumerate}
\item Noise effective bandwidths are consistent with RCA test
campaign results including both white noise and swept source
derived values. \item $1/f$ knee measurements are within
specification ($<50$ mHz). \item White noise, system
temperature and effective bandwidth are consistent with Equation
\ref{e:radiometereq}. \item The sky and reference load temperatures
were not kept lower than 19 K, a deviation from flight conditions.
\item For these input conditions, some channels show some gain
compression.
\end{enumerate}

\section{Summary and discussion}\label{SectionSummary}

The principal results from the RAA campaign are summarized in Table \ref{NPS}. In Section \ref{SectionProcedure}, we discuss how these results were obtained.

\subsection{Procedure}\label{SectionProcedure}

\begin{figure}
\centering
\includegraphics[width=9cm]{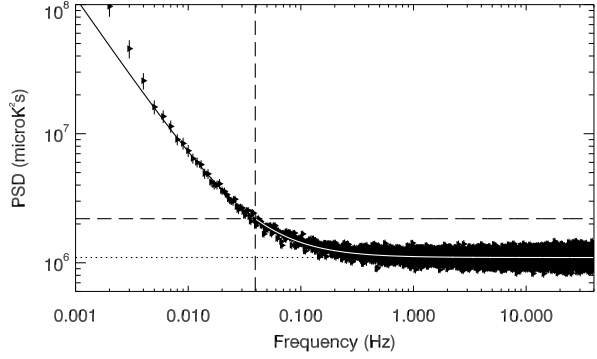}
\caption{Example of a power spectral density and a fit
provided by the model comparison using differenced data from a 70 GHz LFI
diode radiometer (diode $22m1$, as given in Table \ref{NPS}), integrated on the RAA, and viewing a microwave
absorber kept at 22 K. The dotted line shows white noise level. The dashed lines intersect each other at the $1/f$ knee frequency.} \label{FigFit2spectra70}
\end{figure}

For each LFI diode, uncalibrated and calibrated differenced time
streams were produced applying Equations \ref{e:equation_diff} and
\ref{e:equation_diff_calibrated} to the long data set mentioned in Section \ref{SectionRAAResults}. A time stream containing
$10^{5}$ seconds of calibrated differenced data were analyzed in
$25$ individual $4000$ second sections, providing statistics for
estimating the scatter in the noise parameters and a way to weight
the parameter fits. For each data section, a Power Spectral
Density (PSD) was computed. The PSD's for all the sections for
each diode were averaged and white noise, $1/f$ knee and slope
estimated from a fit to Equation \ref{e:PSD}. Additionally, noise
effective bandwidths were estimated from Equation \ref{e:NEB} and, when needed, the results of the correction given by Equation \ref{e:NEBcompressed} were included in the
tabulated values of Table \ref{NPS}. The
formal statistical uncertainties in the parameters are all
approximately $1\%$. We should note that a single value of the
gain modulation factor $r$ was used for the entire $10^{5}$
seconds for each diode, however, no significant change in
parameters was found when $r$ was calculated for the individual
sections. As an example, Figure \ref{FigFit2spectra70} shows a comparison between
data and model for one LFI diode radiometer.

\subsubsection{Noise stability and crosstalk}\label{SectionGMFdiscussion}

These data sets display quite good stability of the noise
parameters. As can be seen in Table \ref{NPS}, standard deviations for white noise and $1/f$ knee are
typically less than $1\%$.

We have also tested a noise model with two independent $1/f$
components added to white noise. The idea behind this model is to try
to separate intrinsic $1/f$ noise, which comes from the
amplifiers, from $1/f$ noise coming from fluctuations in the
temperature of the array or cold loads.  For some data sets, the
two $1/f$ component model gives a slightly better fit, but in
general the single component model fits very well, particularly in
the primary Planck data band (from the spin rate near 0.01 Hz to
the Nyquist sampling rate).

\begin{figure}
\centering
\includegraphics[width=9cm]{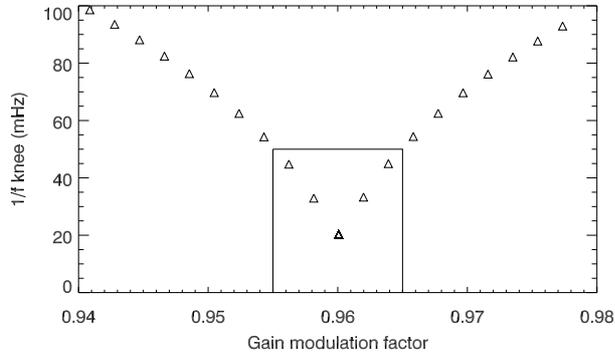}
\caption{Example of the $1/f$ knee
frequency $f_{\mathrm{k}}$ from differenced data as a function of the gain modulation factor $r$. This result was computed using a time stream containing 32 hours of data from a 70 GHz LFI receiver (diode $20m0$). The internal box delimits the region which is within the Planck specification for $1/f$ performance. In this case, we verify that $f_{\mathrm{k}}<50\mathrm{mHz}$ for up to $\pm0.5\%$ variations of the optimal value of $r$.} \label{FigGMF}
\end{figure}

The determination of the gain modulation factor $r$ provides another
test for the robustness of the LFI receivers.  The factor $r$ may be calculated over very long
(month or more) timescales, or over times as short as a single satellite repointing (of order
1 hour), which is the baseline.  We varied the timescale over which $r$ was
calculated for the test data set from 1 to 30 hours, the maximum available. We
find no significant change in white noise or $1/f$ performance.
There is a clear increase in the $1/f$ noise when the $r$ factor
is explicitly set wrong. Figure \ref{FigGMF} shows an example of the dependence of $1/f$ performance as a function of variations in $r$.

The design of LFI is well optimized against crosstalk. Every
detector diode has its own ADC, and the biases are independently
controlled. We have attempted to find an intrinsic cross
correlation among channels with no success. The long data set
discussed here includes a small drift in the temperature of the
load ($0.7$ mK/hour), which dominates any intrinsic cross
correlation in the RCA outputs.

\subsubsection{Frequency spikes}\label{SectionFrequencySpikes}

\begin{figure}
\centering
\includegraphics[width=9cm]{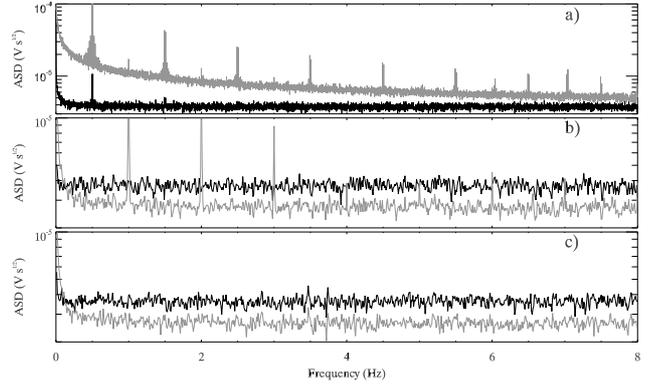}
\caption{Comparison of frequency spikes for different configurations from a 44 GHz LFI receiver (diode $24s1$). ASDs from differenced data are displayed in black. ASDs from sky-only data are displayed in gray. Panel (a) shows data from the RAA test campaign. Note that the frequency spikes start at 0.5 Hz, and are greatly reduced
by differencing. Panels (b) and (c) show data from the functionality test campaign at ambient temperature in Cannes, France. Here the spikes exactly cancel in the difference. Panel (c) shows
data taken in Cannes with the housekeeping acquisition disabled, and clearly shows no spikes at all.} \label{Spikes}
\end{figure}

Some of the LFI receivers exhibit a small artifact, visible in the
power spectra over long periods. The effect is noticeable as a set
of extremely narrow spikes at 1 Hz and harmonics.
These artifacts are nearly identical in sky and reference
samples, and are (almost) completely removed by the LFI
differencing scheme as can be seen in the top panel of Figure \ref{Spikes}.

\begin{figure}
\centering
\includegraphics[width=9cm]{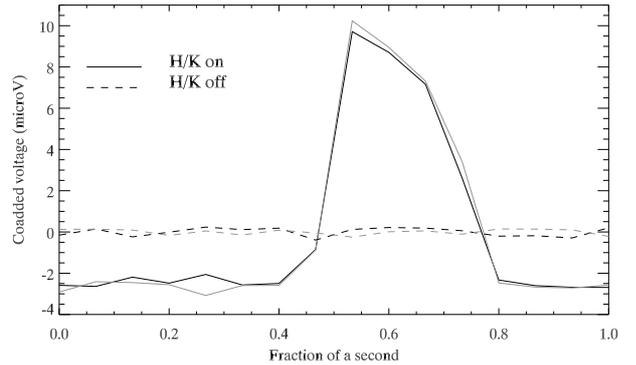}
\caption{Here, one hour of sky-only data from a 44 GHz LFI receiver (diode $24s1$), acquired during the Cannes campaign, have been binned in time, synchronously with the housekeeping acquisition.  We have removed offsets for every second of data for clarity. The disturbance due to the housekeeping acquisition is very evident from about $0.5$ to $0.75$ seconds. Once again we have included data with and without the housekeeping
acquisition enabled.} \label{Bin}
\end{figure}

Extensive testing and analysis has identified
the spikes as a subtle disturbance on the science channels from the
housekeeping data acquisition, which is also performed by the DAE
(albeit with independent ADCs and electronics) at 1 Hz sampling. This causes the
disturbance to be exactly synchronized with the science data, which
makes it more visible and easier to remove. Figure \ref{Spikes} shows data from ambient temperature
functional tests with and without the housekeeping data acquisition operational. These data come from tests done at the satellite
integration level in Cannes, France in 2008.  During these tests the LFI front end amplifiers were in a low gain state, making it easier
to investigate subtle electronic interference such as these spikes.  There are
three significant things to notice here: the spikes only occur when the housekeeping acquisition is
active; the spikes are exactly common mode for the balanced situation of the ambient tests; the spikes
for the Cannes tests are 1 Hz and harmonics.  The earlier test shown in the upper panel included
spikes at harmonics of 0.5 Hz. This has been shown to be an artifact of the RAA test chamber:  all subsequent
tests, including fully integrated satellite tests done in Liege, Belgium in summer 2008 have shown the
well understood 1 Hz spikes. Figure \ref{Bin} demonstrates the way this disturbance is synchronized in time. These data
were taken with very low thermal noise, to enhance the appearance of the disturbance. The data have been
binned synchronously with the one second housekeeping sampling, and the disturbance due to the acquisition is very
clear. This plot is for undifferenced data, the scientific data after difference show no significant disturbance.
Despite the amelioration of the spikes by differencing, software tools have been developed to remove the disturbance from the
limited number of channels showing it, and have been tested on the full Planck system
tests carried out at the Centre Spatial de Li\`{e}ge (CSL), in Belgium, in July and August of 2008.  Part of the
commissioning phase of Planck will include careful on-orbit
characterization of the spikes to further optimize the tools. Monte Carlo testing of
the LFI analysis pipeline includes simulations and removal of
these spikes.

\subsection{Estimated flight sensitivity}\label{SectionEstimatedSensitivity}

The in-flight radiometer's sensitivity was estimated from extrapolating white noise RAA measurements,
obtained at $20$ K sky load temperature, to the expected calibrated sensitivity in
flight conditions. The procedure considers a general radiometric
output model, including non linearity, in which the LFI
receiver voltage output $V_{\mathrm{out}}$ is provided by
\begin{equation}\label{e:fs2}
V_{\mathrm{out}}=\frac{G(T_{\mathrm{noise}}+T_{\mathrm{target}})}{1+bG(T\mathrm{_{noise}}+T_{\mathrm{target}})},
\end{equation}
where $G$ is the photometric calibration in the limit of
linear response, $T_{\mathrm{noise}}$ is the system noise
temperature, $T_{\mathrm{target}}$ is an input temperature, and $b$ is
a non linearity parameter (Daywitt \cite{daywitt89}). These parameters were obtained from dedicated tests during the RAA campaign combined with compression test results from the RCA campaign. We extrapolate the uncalibrated white noise
measured with a sky load temperature of
$20$ K, to estimate calibrated white noise when $T_{\mathrm{target}}$ corresponds to the antenna temperature of the microwave sky. The extrapolation is dominated by the change of the sky load temperature, but the calculation includes a correction for system temperature with FPU temperature, as well as proper noise weighted averaging of the two detector diodes of each radiometer. The LFI sensitivity predictions are provided in Table \ref{FES}.

\section{Conclusion}\label{SectionConclusion}
The Planck-LFI noise has been extensively characterized during several cryogenic
test campaigns. The receivers display exceptional $1/f$ noise performance and
stability, and the estimated sensitivities are within twice the goal values. Careful examination of noise performance results from independent tests at various integration levels has allowed quantitative confirmation of the most important instrumental effects, including compression, noise effective bandwidth, gain modulation factor, and noise artifacts.

   \begin{table}
      \caption[]{Noise performance summary from RAA tests. For convenience, $T_{\mathrm{noise}}$ measurements from RCA tests (Villa \cite{villa09}) are also provided. Temperatures are quoted in Rayleigh-Jeans units.}
         \label{NPS}
     $$
         \begin{array}{lcccccc}
            \hline
            \noalign{\smallskip}
            Diode & \sigma_{T}^{20\mathrm{K}} (\mu\mathrm{K}\sqrt{\mathrm{s}}) & f_{\mathrm{k}} (\mathrm{mHz}) & \alpha & \beta_{\mathrm{eff}} (\mathrm{GHz}) & \chi^{2}_{\nu} & T_{\mathrm{noise}} (\mathrm{K}) \\
            \noalign{\smallskip}
            \hline
            \noalign{\smallskip}
\underline{\mathbf{70 GHz}}& & & & & & \\
18m0 &   -\ (-)  &  -\ (-)  &    -\ (-)  &      - & -  & 36.0 \\
18m1 &   -\ (-)  &  -\ (-)  &    -\ (-)  &      - & -  & 36.1 \\
18s0 & 1124\ (0.02\%) & 61\ (1\%) &     -1.12\ (1\%) &      11.8 & 1.40  & 33.9 \\
18s1 & 1072\ (0.02\%) & 59\ (1\%) &     -1.12\ (1\%) &      15.0 & 1.40  & 35.1 \\
19m0 & 1214\ (0.02\%) & 25\ (2\%) &     -1.27\ (2\%) &      10.1 & 1.26  & 33.1 \\
19m1 & 1165\ (0.02\%) & 32\ (1\%) &     -1.22\ (1\%) &      10.4 & 1.27  & 31.5 \\
19s0 & 1113\ (0.02\%) & 27\ (2\%) &     -1.11\ (2\%) &      10.7 & 1.29  & 32.2 \\
19s1 & 1109\ (0.02\%) & 37\ (2\%) &     -1.02\ (1\%) &      12.1 & 1.28  & 33.6 \\
20m0 & 1094\ (0.02\%) & 21\ (2\%) &     -1.47\ (2\%) &      11.6 & 1.29  & 35.2 \\
20m1 & 1138\ (0.02\%) & 19\ (2\%) &     -1.64\ (3\%) &      10.5 & 1.32  & 34.2 \\
20s0 & 1195\ (0.02\%) & 23\ (2\%) &     -1.27\ (2\%) &      10.6 & 1.31  & 36.9 \\
20s1 & 1145\ (0.02\%) & 28\ (2\%) &     -1.24\ (2\%) &      11.7 & 1.31  & 35.0 \\
21m0 &  866\ (0.02\%) & 28\ (2\%) &     -1.48\ (2\%) &      12.3 & 1.34  & 27.3 \\
21m1 &  891\ (0.02\%) & 30\ (1\%) &     -1.61\ (2\%) &      12.8 & 1.35  & 28.4 \\
21s0 & 1193\ (0.02\%) & 41\ (1\%) &     -1.15\ (1\%) &      12.2 & 1.30  & 34.4 \\
21s1 & 1279\ (0.02\%) & 38\ (1\%) &     -1.17\ (1\%) &      10.8 & 1.29  & 36.4 \\
22m0 & 1029\ (0.02\%) & 46\ (1\%) &     -1.18\ (1\%) &      12.2 & 1.29  & 30.9 \\
22m1 & 1048\ (0.02\%) & 39\ (1\%) &     -1.26\ (1\%) &      11.5 & 1.28  & 30.3 \\
22s0 &  943\ (0.02\%) & 41\ (1\%) &     -1.19\ (1\%) &      13.0 & 1.28  & 30.3 \\
22s1 & 1008\ (0.02\%) & 76\ (1\%) &     -1.01\ (1\%) &      13.6 & 1.30  & 31.8 \\
23m0 & 1038\ (0.02\%) & 30\ (2\%) &     -1.11\ (2\%) &      12.7 & 1.28  & 35.9 \\
23m1 &  964\ (0.02\%) & 31\ (2\%) &     -1.19\ (2\%) &      14.3 & 1.27  & 34.1 \\
23s0 & 1137\ (0.02\%) & 58\ (1\%) &     -1.15\ (1\%) &      13.8 & 1.26  & 33.9 \\
23s1 & 1116\ (0.02\%) & 75\ (1\%) &     -1.12\ (1\%) &      13.5 & 1.28  & 31.1 \\
\underline{\mathbf{44 GHz}}& & & & &  & \\
24m0 &   -\ (-)  &  -\ (-)  &    -\ (-)  &      - & -  & 15.5 \\
24m1 &   -\ (-)  &  -\ (-)  &    -\ (-)  &      - & -  & 15.3 \\
24s0 & 1320\ (0.03\%) & 39\ (1\%) &     -1.06\ (1\%) &      3.6 & 1.27  & 15.8 \\
24s1 & 1129\ (0.03\%) & 46\ (1\%) &     -1.11\ (1\%) &      4.7 & 1.37  & 15.8 \\
25m0 & 1295\ (0.03\%) & 31\ (2\%) &     -1.07\ (1\%) &      4.1 & 1.31  & 17.5 \\
25m1 & 1276\ (0.03\%) & 31\ (2\%) &     -1.03\ (1\%) &      4.2 & 1.45  & 17.9 \\
25s0 & 1280\ (0.03\%) & 21\ (2\%) &     -1.10\ (2\%) &      3.6 & 1.34  & 18.6 \\
25s1 & 1124\ (0.04\%) & 30\ (3\%) &     -1.00\ (2\%) &      4.8 & 2.41  & 18.4 \\
26m0 & 1216\ (0.03\%) & 61\ (1\%) &     -1.01\ (1\%) &      4.3 & 1.31  & 18.4 \\
26m1 & 1257\ (0.03\%) & 61\ (1\%) &     -1.01\ (1\%) &      3.9 & 1.31  & 17.4 \\
26s0 & 1277\ (0.03\%) & 61\ (1\%) &     -1.05\ (1\%) &      3.4 & 1.27  & 16.8 \\
26s1 & 1111\ (0.02\%)  &  -\ (-)  &    -\ (-)  &      4.8 & -  & 16.5 \\
\underline{\mathbf{30 GHz}}& & & & &  & \\
27m0 &  942\ (0.03\%) & 30\ (2\%) &     -1.06\ (2\%) &     4.7 & 1.30  & 12.1 \\
27m1 &  949\ (0.03\%) & 30\ (2\%) &     -1.13\ (2\%) &     4.6 & 1.31  & 11.9 \\
27s0 &  907\ (0.03\%) & 27\ (2\%) &     -1.25\ (2\%) &     5.3 & 1.28  & 13.0 \\
27s1 &  966\ (0.03\%) & 26\ (2\%) &     -1.13\ (2\%) &     4.5 & 1.28  & 12.5 \\
28m0 & 1001\ (0.04\%) & 37\ (2\%) &    -0.94\ (1\%) &      4.1 & 1.29  & 10.6 \\
28m1 & 1051\ (0.04\%) & 31\ (2\%) &    -0.93\ (2\%) &      3.8 & 1.30  & 10.3 \\
28s0 &  917\ (0.03\%) & 37\ (2\%) &    -1.07\ (1\%) &      4.7 & 1.28  & 9.9 \\
28s1 &  929\ (0.03\%) &39\ (2\%) &     -1.06\ (1\%) &      4.6 & 1.29  & 9.8 \\
            \noalign{\smallskip}
            \hline
         \end{array}
     $$
The provided parameters are defined within the model given by Equation \ref{e:PSD}. The errors reported in parentheses are derived from the scatter in the power spectra of individual
data sections (see Section \ref{SectionProcedure}). The LFI radiometers at 70 GHz are identified with RCA labels from 18 to 23. The radiometers at 44
GHz are designated with labels from 24 to 26. The radiometers at
30 GHz are labelled as 27 and 28. The letters $m$ and $s$,
respectively, indicate if a given radiometer is connected to the
main or side OMT. The indexes 0 and 1 identify one of the two
radiometer's diode. RCA 18$m$ and RCA 24$m$ were not operational during RAA testing, and were subsequently repaired. A wrong set of REBA compression parameters was applied to RCA 26$s$1 during the long integration test, and white noise only has been estimated from another (much shorter) data set.
These channels were characterized during the final cryogenic test in July, 2008.
\end{table}

   \begin{table}
      \caption[]{Estimated flight sensitivity from noise measurements extrapolation (see Section \ref{SectionEstimatedSensitivity}). The sensitivity goals for individual radiometers at 30, 44 and 70 GHz were 170, 200, and 270 $\mu\mathrm{K}\sqrt{\mathrm{s}}$, respectively. Requirements to achieve the core scientific aims of LFI are considered to be a factor of two worse than these goals. Estimations are quoted in Rayleigh-Jeans units.}
         \label{FES}
     $$
         \begin{array}{lclc}
            \hline
            \noalign{\smallskip}
            Radiometer &  \sigma_{T}^{flight} (\mu\mathrm{K}\sqrt{\mathrm{s}}) &  Radiometer &  \sigma_{T}^{flight} (\mu\mathrm{K}\sqrt{\mathrm{s}})\\
            \noalign{\smallskip}
            \hline
            \noalign{\smallskip}
\mathbf{70 GHz} & &\mathbf{44 GHz} & \\
\overline{\underline{2\times\mathrm{goal}}} &  \overline{\underline{\mathbf{540}}}  & \overline{\underline{2\times\mathrm{goal}}} & \overline{\underline{\mathbf{400}}}  \\
18m &  -  & 24m &  -  \\
18s & 468 & 24s & 447 \\
19m & 546 & 25m & 501 \\	
19s & 522 & 25s & 492 \\
20m & 574 & 26m & 398 \\		
20s & 593 & 26s & 392 \\
21m & 424 & \overline{\underline{\mathrm{weighted\ mean}}} & \overline{\underline{\mathbf{439}}}  \\	
21s & 530 & \mathbf{30 GHz} & \\
22m & 454 & \overline{\underline{2\times\mathrm{goal}}} & \overline{\underline{\mathbf{340}}} \\		
22s & 463 & 27m & 241 \\
23m & 502 & 27s & 288 \\
23s & 635 & 28m & 315 \\	
\overline{\underline{\mathrm{weighted\ mean}}} &  \overline{\underline{\mathbf{508}}}  & 28s & 251 \\
  &   & \overline{\underline{\mathrm{weighted\ mean}}} & \overline{\underline{\mathbf{269}}}  \\	
\noalign{\smallskip}
            \hline
         \end{array}
     $$
   \end{table}

\begin{figure}
\centering
\includegraphics[width=9cm]{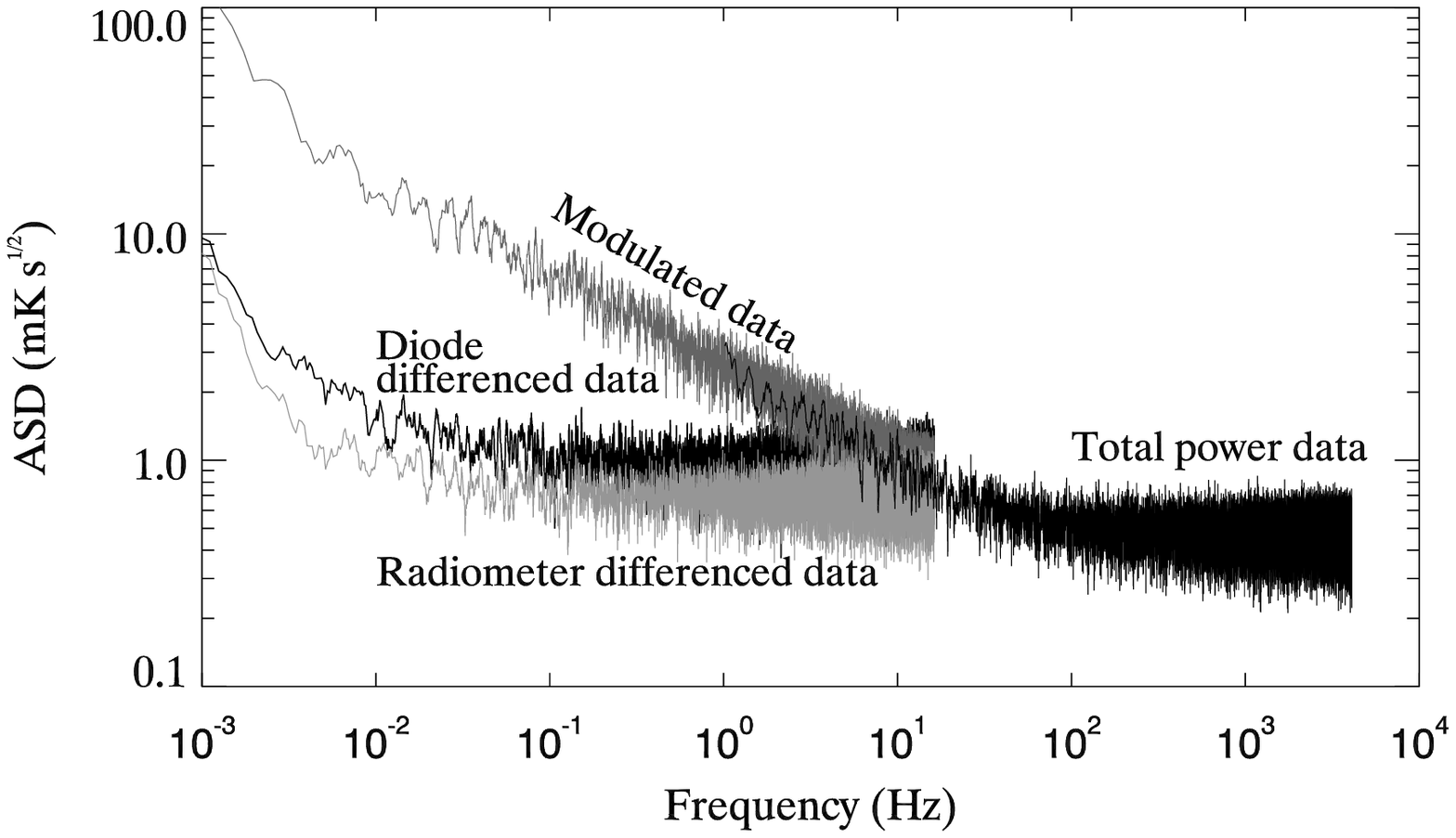}
\caption{Example of an ASD comparison between different data
acquisition modes. The effect of reducing $1/f$ noise due to
switching is self evident. The effect of improving sensitivity by
averaging data from two diodes is also evident. One can also note
the change in the Nyquist frequency due to downsampling. Each
acquisition mode sensitivity is consistent with what was expected
from the most basic model (as described in Appendix A). This example
shows the internal consistency presented by LFI among dedicated
tests.} \label{FigSpectra}
\end{figure}

\begin{appendix}
\section{Receiver sensitivity constant}

In this work, we use Equation \ref{e:radiometereq} to evaluate if, for any given LFI receiver,
white noise corresponds to expectation. In that equation, the
constant $K$ accounts for different radiometer topologies and
differencing techniques. The purpose of this note is to clarify
the relevant $K$ values for LFI, within the model given by
Equation \ref{e:radiometereq}.

\begin{description}
\item[$K=1$.] This is the constant to be applied to Equation
\ref{e:radiometereq} when estimating sensitivity for a single LFI
diode acquiring data in total power mode (i.e. an LFI diode
receiver when not switching). \item[$K=\sqrt{2}$.] This is the
constant to be applied to Equation \ref{e:radiometereq} when
estimating sensitivity for a single LFI diode acquiring data in
modulated mode (i.e. an LFI diode receiver in switched condition).
In this situation, the noise is higher because LFI spends only
half of the available integration time looking to a given target.
This is also the constant to be applied when estimating
sensitivity for a single LFI radiometer (a single LFI radiometer
provides data by averaging differenced data from two independent
and complementary LFI diodes). \item[$K=2$.] This is the constant
to be applied to Equation \ref{e:radiometereq} when estimating
sensitivity for differenced data from a single LFI diode. A
degradation in the noise occurs because we use only half of the
available integration time, and we also add the sky and reference
noise in quadrature when performing the difference described by
Equation \ref{e:equation_diff}.
\end{description}

Figure \ref{FigSpectra} shows a comparison between different
LFI data acquisition modes that clearly illustrates the different
LFI constant sensitivities.

\end{appendix}

\begin{appendix}
\section{Amplitude spectral density normalization}

The amplitude spectral density (ASD) is the square root of the
power spectral density, and it is usually given in units of
$\mathrm{K}/\sqrt{\mathrm{Hz}}$. The $1/\sqrt{\mathrm{Hz}}$
denotes that the value is per unit bandwidth, and is thus
independent of the resolution bandwidth used to compute a result.
Despite the apparent units, $\mathrm{K}/\sqrt{\mathrm{Hz}}$ and
$\mathrm{K}\sqrt{\mathrm{s}}$ are not equivalent, and the purpose
of this note is is to clarify the difference.
\begin{itemize}
\item $\mathrm{K}/\sqrt{\mathrm{Hz}}$ refers to an `integration
bandwidth' of 1 Hz, and assumes by convention a 6 dB/octave
rolloff (obtainable from a 1 pole RC filter). This is the standard
convention for ASD plots for historical reasons and comparisons
with hardware FFT analyzers. \item $\mathrm{K}\sqrt{\mathrm{s}}$
refers to an integration time of 1 second. The effective
integration time $\tau$ of a 1 Hz bandwidth is $0.5$ seconds.
These units are easier for estimating sensitivity versus
integration time.
\end{itemize}
Given these two definitions, we need to keep in mind the following
unit conversion
\begin{equation}\label{e:unitconversion}
    \mathrm{K}/\sqrt{\mathrm{Hz}}=\sqrt{2}\times\mathrm{K}\sqrt{\mathrm{s}}.
\end{equation}
For example, a time stream with white noise only, and samples at 1
second spacing with 1 K RMS, should produce an ASD with $1.414$
$\mathrm{K}/\sqrt{\mathrm{Hz}}$ everywhere. The assumption is that
each sample in the time-ordered data was a $100\%$ duty cycle
integration (i.e. 1 second long integration).
\end{appendix}

\begin{acknowledgements}

Planck is a project of the European Space Agency with instruments
funded by ESA member states, and with special contributions from Denmark
and NASA (USA). The Planck-LFI project is developed by an International
Consortium lead by Italy and involving Canada, Finland, Germany, Norway,
Spain, Switzerland, UK, USA. The US Planck Project is supported by the NASA Science Mission Directorate. The Italian contribution to Planck is supported by ASI - Agenzia Spaziale Italiana. Part of this work was supported by Plan Nacional de I+D, Ministerio de Educaci\'{o}n y Ciencia, Spain, grant reference ESP2004-07067-C03-02. TP's work was supported in part by the Academy of Finland grants 205800, 214598, 121703, and 121962. TP thanks Waldemar von Frenckells stiftelse, Magnus Ehrnrooth Foundation, and V\"ais\"al\"a Foundation for financial support. We acknowledge the use of the Lfi Integrated perFormance Evaluator (LIFE) package (Tomasi \cite{tomasi09}).
\end{acknowledgements}

\end{document}